\documentclass[ twocolumn,08 pt, amsmath,amssymb, aps,fleqn]{revtex4-1}
\usepackage[hidelinks]{hyperref}
\usepackage[english]{babel}
\usepackage{mathtools}
\usepackage{tabularx}
\usepackage{multirow}
\usepackage{comment}
\usepackage{graphicx}
\usepackage{color}
\usepackage{soul}
\begin{document}
\preprint{APS/123-QED}
\title{Friedel oscillations and dynamical density of states of an inhomogeneous Luttinger liquid}
\author{Joy Prakash Das}  \author{Chandramouli Chowdhury} \author{Girish S. Setlur}\email{gsetlur@iitg.ernet.in}
\affiliation{Department of Physics \\ Indian Institute of Technology  Guwahati \\ Guwahati, Assam 781039, India}

\begin{abstract}
In this work, the four-point Green functions relevant to the study of Friedel oscillations are calculated for a Luttinger liquid with a cluster of impurities around an origin using the powerful Non chiral bosonization technique (NCBT). The two-point functions obtained using the same method are used to calculate the dynamical density of states (DDOS), which exhibits a power law in energy and closed analytical expressions for the DDOS exponent is calculated. These results interpolates between the weak barrier and weak link cases which are typically studied in the literature. The dependence of the DDOS on the nature of interactions and the strength of the impurity clusters are highlighted.  Finally the special case of the Luttinger parameter g=1/2 is studied and compared with existing results.
\end{abstract}

\maketitle
\section{Introduction}

One dimensional systems hold a special position in the study of many body physics due to the unusual nature of the mutual interactions between the constituent particles, which forbids the use of Fermi liquid theory and perturbative techniques. An alternative is invoked to describe the state of 1D interacting systems, which goes by the name `Luttinger liquid' \cite{haldane1981luttinger} which is based on the linearization of the dispersion relations near to the Fermi level. The study of interactions is quantified by the computation of the N-point correlation functions using which a number of physical phenomena can be described. The most prevalent method  to calculate the correlation functions of one dimensional systems is bosonization, which expresses a fermionic field as the exponential of a bosonic field \cite{von1998bosonization}. This field theoretical approach to bosonization, which goes under the name g-ology is well established \cite{giamarchi2004quantum} and can successfully compute the N-point Green functions of a homogeneous Luttinger liquid. But the introduction of an impurity leads to inadequacy of this method and to circumvent this, other techniques like renormalization group (RG) methods are called for \cite{matveev1993tunneling}. 

More recently, a new technique has been developed which can extract the most singular part of the correlation functions of a Luttinger liquid with arbitrary strength of the external impurities as well as that of mutual interactions between the particles \cite{das2018quantum}. This method, which goes by the name `Non chiral bosonization technique'  has been applied successfully to study the one step fermionic ladder (two 1D wires placed parallel and close to each other with hopping between a pair of opposing points) \cite{das2017one} and slowly moving heavy impurities in a Luttinger liquid \cite{das2018ponderous}. Once the Green function of a given system is obtained, it may be used to predict different physical phenomena occurring in the system. Typical physical phenomena studied in Luttinger liquids with impurities includes Friedel oscillations \cite{Egger1995friedel1}, conductance \cite{fendley1995exact, fendley1995exact2}, Kondo effect \cite{furusaki1994kondo, schiller1995exact} and so on. In the famous work by Kane and Fisher \cite{kane1992transport}, it has been shown how impurities can bring drastic effects to the conductance of the particles based on their nature of interaction, viz., attractive or repulsive. It is also well known that non trivial interactions between electrons and impurities leads to the occurrence of Friedel oscillations in the charge density profile of Fermi liquids  \cite{friedel1958metallic, tutto1988theory, simion2005friedel}.  

In Luttinger liquids too, it is quite interesting to study the interplay between mutual interactions and impurities by taking into account Friedel oscillations. Egger and Grabert have studied Friedel oscillations in a Luttinger liquid with arbitrary interactions and arbitrary strengths of impurities  \cite{Egger1995friedel1} by combining the techniques of standard bosonization \cite{emery1979highly, haldane1981luttinger, kane1992transport}, self-consistent harmonic approximation \cite{saito1978self, gogolin1993local} and quantum Monte Carlo simulations. Fernandez et al. provided an alternative approach to this problem based on a path-integral bosonization technique previously developed in the context of non local quantum field theories and specially suited to consider long-ranged interactions \cite{fernandez2001friedel}. Grishin et al. studied Friedel oscillations in a Luttinger liquid using the Green functions obtained using functional bosonization \cite{grishin2004functional}. They also described the suppression of electron local density of states (LDOS) at the position of the impurity and obtained an analytic expression for the LDOS at any distance from the impurity.  For the spinless Luttinger liquid and strong impurity, the density of states has been exactly calculated by Von Delft et.al. \cite{von1998bosonization} for the specific case of $K_{\rho}=0.5$ where $K_{\rho}$ is the Luttinger parameter. With the emergence of 1D materials like carbon nano-tubes \cite{bockrath1999luttinger} and quantum wires \cite{auslaender2002tunneling}, the theoretical studies can be expected to make contact with experiments. 

In this work, the NCBT is used, which uses a non standard harmonic analysis to express the fast parts of the density correlation functions in terms of the slowly varying parts which is suitable for studying the systems under consideration. The oscillating terms in the density density correlation functions (DDCF) are then calculated as a special case of four point functions. The two-point correlation functions for the same class of systems are used to calculate the DDOS, which for a Luttinger liquid exhibits a power law in energy and closed analytical expressions for the exponent is obtained for the same both for points near to and far away from the impurity. A comparison with existing literature is also made. The highlight of this work is that while these physical phenomena are studied using combination of different techniques besides bosonization, NCBT tackles these issues in an easier way.

\section{System description}

The system consists of a one dimensional chain of electrons with short ranged mutual interactions in the presence of a finite number of barriers and wells clustered around an origin. The Hamiltonian of the system is given as follows.
\small
\begin{equation}
\begin{aligned}
H =& \int^{\infty}_{-\infty} dx \mbox{    } \psi^{\dagger}(x) \left( - \frac{1}{2m} \partial_x^2 + V(x) \right) \psi(x)\\
  & \hspace{1cm} + \frac{1}{2} \int^{ \infty}_{-\infty} dx \int^{\infty}_{-\infty} dx^{'} \mbox{  }v(x-x^{'}) \mbox{   }
 \rho(x) \rho(x^{'})
\label{Hamiltonian}
\end{aligned}
\end{equation}
\normalsize
The first two terms are the kinetic and potential energy term respectively. The potential cluster can consist of one delta impurity $V_0\delta(x)$, two delta impurities placed close to each other $V_0( \delta(x+a)+\delta(x-a))$, finite barrier/well $\pm V \theta(x+a)\theta(a-x)$, etc., where $\theta(x)$ is the Heaviside step function. The system is subjected to the RPA (random phase approximation) which enables the calculation of the analytical expressions of the correlation functions. In this limit, the Fermi momentum and the mass of the fermion are allowed diverge in such a way that their ratio is finite (i.e. $ k_F, m \rightarrow \infty $ but $ k_F/m = v_F < \infty  $). Units are chosen such that $ \hbar = 1 $ and hence $ k_F $ is both the Fermi momentum as well as a wavenumber \cite{stone1994bosonization}. The RPA limit results in linearizing the energy momentum dispersion near the Fermi surface ($E=E_F+p v_F$ instead of $E=p^2/(2m)$). Furthermore, if `2a' is the width of the impurity cluster (distance between the two delta potentials or width of the finite barrier) it is then imperative to define how this width `2a' scales in the RPA limit. The assertion made is that in the RPA limit, $ 2 a k_F   < \infty $ as $ k_F \rightarrow \infty $. In a similar way, the heights and depths of the various barriers/wells are assumed to be in fixed ratios with the Fermi energy $ E_F = \frac{1}{2} m v_F^2 $ even as $ m \rightarrow \infty $ with $ v_F < \infty $. In case of the different potentials consisting the cluster, the only quantities that will be used in the calculation of the Green functions is the reflection (R) and transmission (T) amplitudes which can be easily calculated using elementary quantum mechanics and are provided in an earlier work \cite{das2018quantum}. Here the generalized notion of R and T is used in this work to signify the reflection and transmission amplitudes of the cluster of impurities in consideration. The third term in eq. (\ref{Hamiltonian}) is the forward scattering mutual interaction term such that
\begin{equation*} 
v(x-x^{'}) = \frac{1}{L} \sum_{q}  v_q \mbox{ }e^{ -i q(x-x^{'}) } 
\end{equation*}
where $ v_q = 0 $ if $ |q| > \Lambda $ for some fixed bandwidth $ \Lambda \ll k_F $ and $ v_q = v_0 $ is a constant, otherwise.\\

\section{ Non chiral bosonization and two point functions}
As in conventional bosonization schemes using the field theoretical approach, the fermionic field operator is expressed in terms of currents and densities. But in NCBT the field operator is modified to include the effect of back-scattering by impurities as follows.
\begin{equation}
\begin{aligned}
\psi_{\nu}(x,\sigma,t) \sim C_{\lambda  ,\nu,\gamma}\mbox{ }e^{ i \theta_{\nu}(x,\sigma,t) + 2 \pi i \lambda \nu  \int^{x}_{sgn(x)\infty}\mbox{ } \rho_s(-y,\sigma,t) dy}
\label{PSINU}
\end{aligned}
\end{equation}
Here $\theta_{\nu}$ is the local phase which is a function of the currents and densities which is also present in the standard bosonization schemes \cite{giamarchi2004quantum}, ideally suited for homogeneous systems.
\small
\begin{equation}
\begin{aligned}
\theta_{\nu}(x,\sigma,t) =& \pi \int^{x}_{sgn(x)\infty} dy \bigg( \nu  \mbox{  } \rho_s(y,\sigma,t)\\
&\hspace{1 cm} -  \int^{y}_{sgn(y)\infty} dy^{'} \mbox{ }\partial_{v_F t }  \mbox{ }\rho_s(y^{'},\sigma,t) \bigg)
\end{aligned}
\end{equation}\normalsize
The new addition in eq. (\ref{PSINU}) is the $\rho_s(-y)$ term which ensures the necessary trivial exponents for the single particle Green functions for a system of otherwise free fermions with impurities. The single particle Green functions of free fermions with impurities can be obtained using standard Fermi algebra and they serve as a basis for comparison for the Green functions obtained using the bosonized version of the field operator in eq. (\ref{PSINU}). The adjustable parameter is the quantity $\lambda$ which can be either 0 or 1 as per requirement. Thus the standard bosonization scheme can be easily obtained by setting $\lambda=0$. The factor $2 \pi i$ ensures that the fermion commutation rules are preserved. The quantities $C_{\lambda  ,\nu,\gamma}$ are also fixed by this comparison and they involve pre-factors which do not show any dynamics. The suffix $\nu$ signifies a right mover or a left mover and takes values 1 and -1 respectively. This field operator (annihilation) as given in eq. (\ref{PSINU}), to be treated as a mnemonic to obtain the Green functions and not as an operator identity, is clubbed together with another such field operator (creation) to obtain the non interacting two point functions after fixing the C's and $\lambda$'s. Finally the densities $\rho$'s are replaced by their interacting versions to obtain the many body Green functions, the details being described in an earlier work \cite{das2018quantum}. The two point functions obtained using NCBT are given in \hyperref[AppendixA]{Appendix A}.

\section{Four-point functions (Friedel oscillations)}\normalsize

In the RPA sense, the density $ \rho(x,\sigma,t) $ may be ``harmonically analysed" as follows.
\begin{equation}\label{INPUT2}\footnotesize
\rho(x,\sigma,t) = \rho_s(x,\sigma,t) + e^{ 2 i k_F x } \mbox{   }\rho_f(x,\sigma,t) +  e^{ - 2 i k_F x } \mbox{   }\rho^{*}_f(x,\sigma,t)
\end{equation}\normalsize

Here $ \rho_s $ and $ \rho_f $ are the slowly varying and the rapidly varying parts respectively. The auto-correlation function of the slowly varying part of the density $ \rho_s $ (the average density is subtracted out, so this is really the deviation) may be written down using Wick's theorem to give rise to the density density correlation functions of free fermions as follows.

\footnotesize
\begin{equation}
\begin{aligned}\label{INPUT3}
\langle T \mbox{    }&\rho_s(x_1,\sigma_1,t_1)\rho_s(x_2,\sigma_2,t_2)\rangle_0\\
&=-\delta_{\sigma_1,\sigma_2}\frac{1}{4\pi^2}\sum\limits_{\nu = \pm 1} 
\bigg(\frac{\theta(x_1x_2)}{ [\nu (x_1 - x_2) - v_F (t_1-t_2) ]^2 }\\
&+\frac{|R|^2\mbox{ }\theta(x_1x_2)}{ [\nu (x_1 + x_2) - v_F (t_1-t_2) ]^2 }+\frac{(1-|R|^2)\theta(-x_1x_2)}{ [\nu (x_1 - x_2) - v_F (t_1-t_2) ]^2 }\bigg)\\
\end{aligned}
\end{equation}
\normalsize
Here $|R|^2$ is the reflection coefficient of the cluster of impurities under consideration and $\theta(x)$ is the Heaviside step function. When interactions are taken into account, due to spin charge separation, two different velocities are seen, viz., the holon velocity and the spinon velocity. Thus the density density correlation functions (DDCF) in presence of interactions is given by\small
\begin{equation}
\begin{aligned}\label{DDCFslow}
\langle T \mbox{    }&\rho_s(x_1,\sigma_1,t_1)\rho_s(x_2,\sigma_2,t_2)\rangle \\
=& \frac{1}{4} (\langle T \mbox{    }\rho_h(x_1,t_1)\rho_h(x_2,t_2)\rangle  + \sigma_1 \sigma_2 \langle T \mbox{    }\rho_n(x_1,t_1)\rho_n(x_2,t_2)\rangle )\\
\end{aligned}
\end{equation}
\normalsize
 where $ \rho_h(x,t) =  \rho_s(x,\uparrow,t) + \rho_s(x,\downarrow,t)  $ is the ``holon" density and $ \rho_{n}(x,t) =  \rho_s(x,\uparrow,t) - \rho_{s}(x,\downarrow,t)  $ is the ``spinon" density and
\footnotesize
\begin{equation}
\begin{aligned}
\langle T \rho_a(x_1,t_1)\rho_a(x_2,t_2)\rangle  = \frac{v_F  }{ 2\pi^2 v_a } &\mbox{   } \sum_{  \nu = \pm 1 }\bigg (   \frac{-1}{ ( x_1-x_2 + \nu v_a(t_1-t_2) )^2 }\\
-\frac{ |R|^2 }{    \bigg( 1 - \delta_{a,h} \frac{(v_h-v_F)}{ v_h } |R|^2   \bigg) }&	  \frac{\frac{v_F }{v_a}  \mbox{    } \text{sgn}(x_1) \text{sgn}(x_2)\mbox{   }}{  ( | x_1|+|x_2 | + \nu v_a(t_1-t_2) )^2 }
\bigg)
\label{RHOSRHOS}
\end{aligned}
\end{equation}\normalsize
where $ a = n$ (spinon) or  h (holon). Here the spinon velocity is non-different from the Fermi velocity since it is the total density that couples to the short-range potential ($v_n=v_F$). On the other hand, the holon velocity is modified by interactions, \scriptsize $ v_h = \sqrt{v_F^2+2v_F v_0/\pi} $ \normalsize where $v_0$ is the strength of interaction between fermions as already described in Section 2. 

The slow part of the DDCF given by eq. (\ref{DDCFslow}) can be used to obtain the fast part of the DDCF which corresponds to Friedel oscillations, which is nothing but a term which oscillates with wavenumber $ 2 k_F $ such as   $ e^{ 2 i k_F (x-x^{'}) }\scriptsize <T\mbox{   } \rho_f(x,\sigma,t)\rho^{*}_f(x^{'},\sigma',t^{'})  >$.  This can be done using a non standard harmonic analysis suited to study inhomogeneous Luttinger liquids like the one under study.
\begin{equation}
\hspace{1.5 cm}\rho_f(x,\sigma,t)  \sim e^{ 2 i \pi \int^{x}_{-\infty} (\rho_s(y,\sigma,t) + \lambda \rho_s(-y,\sigma,t)) dy }
\label{RHOFAST}
\end{equation}

The above equation is the basis of the NCBT using which eq. (\ref{PSINU}) is derived. The $\lambda$ is the same as  that of eq. (\ref{PSINU}) taking values 0 or 1 and setting $\lambda = 0$ yields the standard harmonic analysis of Haldane. The value of $\lambda$ is first decided by calculating the non-interacting DDCF using eq. (\ref{RHOFAST}) and comparing with the same DDCF obtained using Fermi algebra. After that, similar to the calculation of the two-point functions, the non interacting DDCF in eq. (\ref{INPUT3}) is to be replaced by the interacting DDCF in eq. (\ref{DDCFslow}) to obtain the required four-point functions in presence of mutual interactions.


Define $ {\tilde{\rho}}_f \equiv \rho_f -<\rho_f> $. The prescription for choosing $ \lambda_i $ in eq. (\ref{RHOFAST}) as discussed in \cite{das2018quantum} leads  to the unambiguous conclusion that $\lambda_1=1-\lambda_2$ (where $\lambda_1$ and $\lambda_2$ corresponds to the points $x_1$ and $x_2$ respectively) and the fast parts of the DDCF corresponding to Friedel oscillations are obtained as follows.,\footnotesize

\begin{equation}
\begin{aligned}
\Big\langle T&\mbox{  } {\tilde{\rho}}_f(x_1,\sigma_1,t_1)  {\tilde{\rho}}_f(x_2,\sigma_2,t_2)\Big\rangle  \sim \\
&  ( Exp[ \sum_{ \substack{\nu,\nu^{'} = \pm 1 \\ a = h,n }   }\Gamma(\nu,\nu^{'};a) \mbox{  } Log[ (\nu x_1 - \nu^{'} x_2 ) - v_a (t_1-t_2) ] ]-1 ) \\
\Big\langle T&\mbox{  } {\tilde{\rho}}_f(x_1,\sigma_1,t_1)  {\tilde{\rho}}^{*}_f(x_2,\sigma_2,t_2)\Big\rangle  \sim  \\
& ( Exp[ - \sum_{ \substack{\nu,\nu^{'} = \pm 1 \\ a = h,n }   }\Gamma(\nu,\nu^{'};a) \mbox{  } Log[ (\nu x_1 - \nu^{'} x_2 ) - v_a (t_1-t_2) ] ]-1 ) \\
\end{aligned}
\end{equation}

\normalsize One should remember that this really means the time derivative of the logarithms of both sides are equal to each other. The values of the anomalous scaling exponents $\Gamma(\nu,\nu^{'};a)$ can be obtained from the expression below.\\
\begin{equation}
\begin{aligned}\footnotesize
\Gamma(\nu,\nu^{'};a) = \left(\frac{v_F}{2v_h}\mbox{ }\delta_{a,h}+\frac{1}{2} \mbox{ }\delta_{a,n} \right)\mbox{ } (\delta_{\nu,\nu'}-\delta_{\nu,-\nu'})
\end{aligned}
\end{equation}

\section{Dynamical density of states}
\label{DDOSCOND}

In this section the results for the local dynamical density of states (DDOS), $ D_x(\omega) $ at location $ x $ is presented. Physically, $ D_x(\omega) d\omega $ is the number of quasiparticle states per unit length with energy between $ \hbar \omega $ and $ \hbar (\omega + d\omega) $ relative to the Fermi energy. In a 1D system of Fermi gas, the density of states is constant for $\omega$ small compared to the Fermi energy and is given by,

\begin{equation}
D(\omega)  =  \frac{1}{ \pi v_F }
\label{METH1}
\end{equation}

For a Fermi liquid also, the density of states near the Fermi energy is constant. But for a Luttinger liquid, the DDOS exhibits a power law in energy: $ D(\omega) \sim \omega^{\alpha} $ where $ \alpha $ is the density of states exponent. $ \alpha $ depends on the forward scattering interaction strength and vanishes when this is zero. To prove this, it is important to generalize the idea of density of states to interacting many body systems. The generalization is given below.

\begin{equation}
D_{ {\bf{x}} }(\omega)  = \int^{\infty}_{-\infty} \frac{dt}{2\pi} \mbox{   }
     e^{ i t ( \omega  + E_F)  } \mbox{  } \langle \{ \psi({\bf{x}},\sigma,t), \psi^{\dagger}({\bf{x}},\sigma,0) \} \rangle
\label{Local}
\end{equation}

The above eq. (\ref{Local}) relates the local density of states to the single particle Green function formulas which are provided in \hyperref[AppendixA]{Appendix A}. The DDOS consists of both slowly and rapidly oscillating (Fridel oscillation) terms which can be written in terms of the different parts of the Green functions as given in eq. (\ref{break}).

\begin{equation*}
\begin{aligned}
D_{ x }(\omega)
 =& \int^{\infty}_{-\infty} \frac{dt}{2\pi} \mbox{   } e^{ i t \omega  }\mbox{  } \Big( \langle \{ \psi_R(x,\sigma,t), \psi^{\dagger}_R(x,\sigma,0) \} \rangle\\
&\hspace{1cm} +  \langle \{ \psi_L(x,\sigma,t), \psi^{\dagger}_L(x,\sigma,0) \} \rangle \Big)\\
+  &\int^{\infty}_{-\infty} \frac{dt}{2\pi} \mbox{   } e^{ i t  \omega  }\mbox{  } \Big( e^{ 2 i k_F x} \mbox{   }\langle \{ \psi_R(x,\sigma,t), \psi^{\dagger}_L(x,\sigma,0) \} \rangle   \\
&\hspace{1cm}+e^{ -2 i k_F x} \mbox{   } \langle \{ \psi_L(x,\sigma,t), \psi^{\dagger}_R(x,\sigma,0) \} \rangle \Big)
\end{aligned}
\end{equation*}

In the present work we shall be content at evaluating the slow contributions to the local DDOS (for fermions with spin, it is calculated for one spin projection and the answer is then doubled). For the inhomogeneous systems under study (cluster of impurities), there are two interesting limits viz. when $ x $ is far away from the cluster of barriers and wells and also when $ x $ is near or at the location of barriers and wells. For this, a dimensionless parameter proportional to the location $ x $ is defined viz. $ \xi = \frac{ 2 \omega |x |}{ v_h } $ and the DDOS at zero temperature is given in general as follows,

\begin{equation}
D_{ \xi }(\omega) \sim \omega^{ \alpha(\xi) }
\label{LDDOS}
\end{equation}

At the Fermi level, local density of states at {\it{finite}}  temperature is obtained by simply replacing $ \omega $ in the above equation by $ k_B T $ viz. the temperature. The exponent $\alpha(\xi)$ is given as follows.
\begin{equation}
\begin{aligned}
\alpha(\xi) = 
\begin{cases}
\frac{(v_h-v_F)^2 }{4 v_F v_h}+\frac{|R|^2(v_h-v_F)(v_h+v_F)}{4  v_h (v_h-|R|^2 (v_h-v_F))} \mbox{ } &0 < \xi \ll 1\\
\frac{(v_h-v_F)^2 }{4 v_F v_h} \mbox{ } &\mbox{ }\mbox{ }\mbox{ }\xi \gg 1 
\end{cases}
\end{aligned}
\label{alpha}
\end{equation}

\subsection{Far away from the impurity}
\begin{figure}[h!]
  \centering
  \includegraphics[scale=0.3]{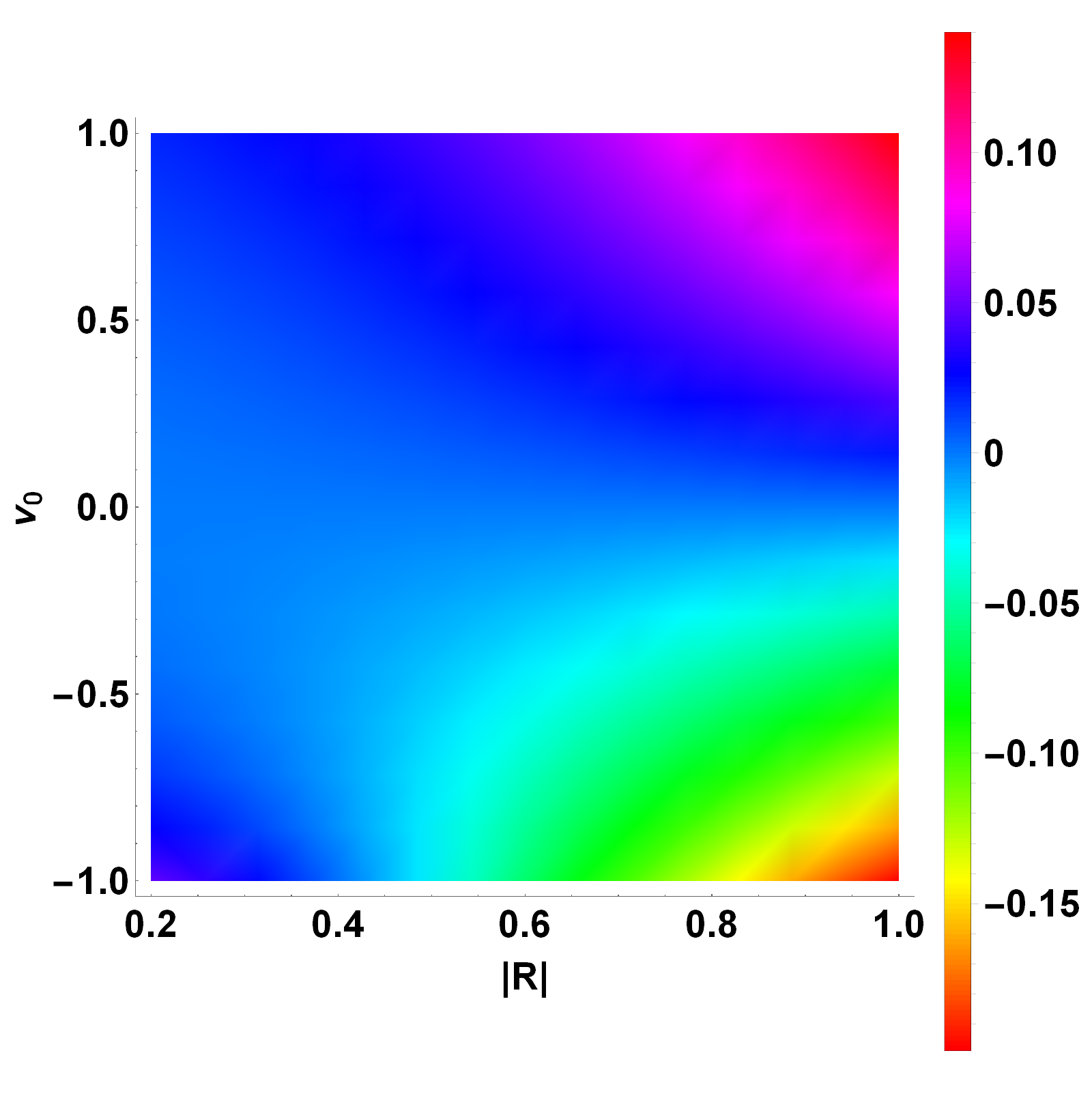}\\(a)\\
\includegraphics[scale=0.3]{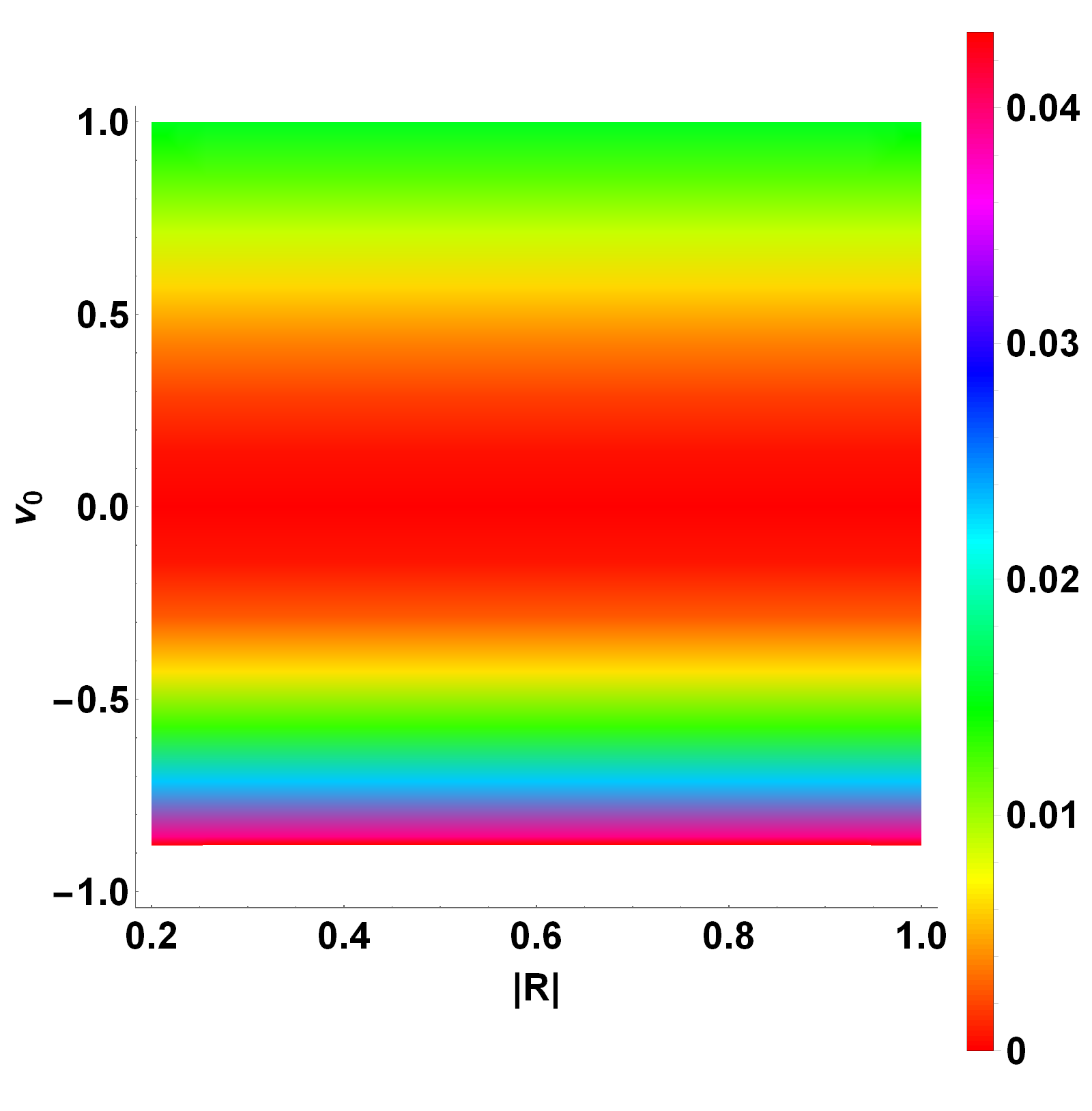}\\(b)
  \caption{(a) Density of state exponents$( 0 < \xi \ll 1 )$ as a function of the interaction parameter $v_0$ for various values of reflection amplitudes $|R|$ given in  parenthesis. (b) Density of state exponents$(  \xi \gg 1 )$ as a function of the interaction parameter $v_0$.($v_F=1$)}\label{DDOSDEN}
\end{figure}

Consider the Luttinger parameter g (also called $K_{\rho}$ in the literature) which in the present work is equal to $v_F/v_h$. From eq. (\ref{alpha}), the DDOS exponent for points far away from the impurity can be written in terms of g  as $\alpha(\xi \gg 1)=\frac{1}{4} (g + \frac{1}{g} - 2 )$  which is precisely the exponent found in the textbooks for fermions with spin (Giamarchi \cite{giamarchi2004quantum}, eq. (7.27)). From fig. \ref{DDOSDEN}. (b), it can be seen that the exponent $\alpha$ is non-negative and its value monotonically increases with an increase in the strength of mutual interactions both for attractive and repulsive interactions. It is independent of the strength of the impurity and it vanishes when interactions are switched off giving back the uniform density of states given by eq. (\ref{METH1}).


\subsection{Near the impurity}

The exponent $\alpha(0 < \xi \ll 1)$ is the new result which is the DDOS exponent for points close to or at the position of the impurity for arbitrary strength of impurities and mutual interactions. It interpolates between the weak impurity and weak link results which are typically discussed in the literature. In the case of the half line, Kane and Fisher \cite{kane1992transport} have remarked that for spinless fermions the density of states is $ \rho_{end}(\epsilon) \sim \epsilon^{ \frac{1}{g} - 1 } $. For fermions with spin the exponent may be inferred as half of this \cite{das2018quantum} viz. $ D_{half-line}(\omega) \sim \omega^{ \frac{1}{2} ( \frac{1}{ g } - 1 ) } $. Setting $ |R| = 1 $ for half-line in eq. (\ref{alpha}), $ \alpha( \xi \equiv 0 ) = \frac{(v_h-v_F) }{2 v_F } = \frac{1}{2} ( \frac{1}{ g} - 1 ) $ since $ g = \frac{ v_F}{v_h} $. Thus the exponents are in exact agreement with that of Kane and Fisher.
 For repulsive interactions,   $ v_h > v_F   $  the exponent $\alpha>0$. This means the density of states near the impurity tends to vanish at low energies for repulsive interactions. This is analogous to the well known phenomena called `cutting the chain' described by Kane and Fisher \cite{kane1992transport}. But for attractive interactions $ v_h < v_F $ and  for $ |R|^2 < |R_{c1}|^2 \equiv \frac{v_F-v_h}{3 v_F-v_h}  $ the exponent $ \alpha > 0 $. Lastly for attractive interactions with $ |R| > |R_{c1}| $ the exponent $ \alpha < 0 $.  The exponent near the impurity becomes negative in some regions signifying that the density of states diverges at low energies near the impurity i.e. the impurity together with attractive interactions effectively brings back low-energy quasiparticles which ought not to be there in a Luttinger liquid.  This is analogous to the well known phenomenon called `healing the chain' \cite{kane1992transport}.

\subsection{Comparison with existing studies (g=1/2 case)}
  For the spinless case and strong impurity, the density of states has been exactly calculated by Von Delft et.al. \cite{von1998bosonization} for the specific case of $g=\frac{1}{2}$ and obtained to be $D(\omega)\sim\omega$. Recalculating the exponent in eq. (\ref{alpha}) for the spinless case (double of this exponent) and strong impurity ($|R|=1$), NCBT yields $D(\omega)\sim\omega^{\alpha}$ where $\alpha = \frac{1}{g} - 1$ and for $g=\frac{1}{2}$, we have $D(\omega)\sim\omega$ which is in agreement with Von Delft et.al. \cite{von1998bosonization}, Fabrizio \& Gogolin\cite{fabrizio1997comment}, Furusaki \cite{furusaki1997local}, etc.  

 The main advancement of the present work is being able to provide simple analytical expressions for exponents such as these that interpolate between the no-barrier and strong barrier cases. The novel technical framework that abandons the g-ology framework in favor of non-chiral bosonization technique with non-standard harmonic analysis of the field operator  enables an exact treatment of free fermions plus impurity problem.

\section{Conclusions}
In this work, the non chiral bosonization technique is used to calculate the rapidly oscillating parts of the density density correlation functions, also called Friedel oscillation terms, which arises because of the presence of a localized impurity in an otherwise homogeneous Luttinger liquid. The two-point functions obtained using the same technique are used to express the local dynamical density of states as a power law and a closed analytical expression of the exponent is obtained as a function of the strength of impurities and that of mutual interactions. The interesting limits of far away from the impurity as well as near or at the position of impurity is also discussed. A comparison is made with the existing literature for the dynamical density of states in the special case of g=1/2 and an exact match is observed.\\\\

\section{APPENDIX A:  Two point functions using NCBT  }
\label{AppendixA}
\setcounter{equation}{0}
\renewcommand{\theequation}{A.\arabic{equation}}

The full Green function is the sum of all the parts. The notion of weak equality is introduced which is denoted by \begin{small} $ A[X_1,X_2] \sim B[X_1,X_2] $ \end{small}. This really means  \begin{small} $ \partial_{t_1} Log[ A[X_1,X_2] ]  = \partial_{t_1} Log[ B[X_1,X_2] ] $\end{small} assuming that A and B do not vanish identically. In addition to this, the finite temperature versions of the formulas below can be obtained by replacing $ Log[Z] $ by $ Log[ \frac{\beta v_F }{\pi}Sinh[ \frac{\pi Z}{ \beta v_F} ] ] $ where $ Z \sim  (\nu x_1 - \nu^{'} x_2 ) - v_a (t_1-t_2)  $ and singular cutoffs ubiquitous in this subject are suppressed in this notation for brevity - they have to be understood to be present. {\bf Notation:} $X_i \equiv (x_i,\sigma_i,t_i)$ and  $\tau_{12} =  t_1 - t_2$. 
\scriptsize

\begin{equation}
\begin{aligned}
\Big\langle T\mbox{  }\psi(X_1)\psi^{\dagger}(X_2) \Big\rangle 
=&\Big\langle T\mbox{  }\psi_{R}(X_1)\psi_{R}^{\dagger}(X_2) \Big\rangle +\Big \langle T\mbox{  }\psi_{L}(X_1)\psi_{L}^{\dagger}(X_2)\Big\rangle \\
+&\Big\langle T\mbox{  }\psi_{R}(X_1)\psi_{L}^{\dagger}(X_2) \Big\rangle + \Big\langle T\mbox{  }\psi_{L}(X_1)\psi_{R}^{\dagger}(X_2)\Big\rangle \\
\label{break}
\end{aligned}
\end{equation}

\small
\begin{bf} Case I : $x_1$ and $x_2$ on the same side of the origin\end{bf} \\ \scriptsize

\begin{equation}
\begin{aligned}
\Big\langle T\mbox{  }\psi&_{R}(X_1)\psi_{R}^{\dagger}(X_2)\Big\rangle \sim 
\frac{(4x_1x_2)^{\gamma_1}}{(x_1-x_2 -v_h \tau_{12})^{P} (-x_1+x_2 -v_h \tau_{12})^{Q}} \\
\times&\frac{1}{ (x_1+x_2 -v_h \tau_{12})^{X} (-x_1-x_2 -v_h \tau_{12})^{X} (x_1-x_2 -v_F \tau_{12})^{0.5}}\\
\Big\langle T\mbox{  }\psi&_{L}(X_1)\psi_{L}^{\dagger}(X_2)\Big\rangle \sim 
\frac{(4x_1x_2)^{\gamma_1}}{(x_1-x_2 -v_h \tau_{12})^{Q} (-x_1+x_2 -v_h \tau_{12})^{P}} \\
\times&\frac{1}{ (x_1+x_2 -v_h \tau_{12})^{X} (-x_1-x_2 -v_h \tau_{12})^{X}(-x_1+x_2 -v_F \tau_{12})^{0.5}}\\
\Big\langle T\mbox{  }\psi&_{R}(X_1)\psi_{L}^{\dagger}(X_2)\Big\rangle \sim 
\frac{(2x_1)^{\gamma_1}(2x_2)^{1+\gamma_2}+(2x_1)^{1+\gamma_2}(2x_2)^{\gamma_1}}{2(x_1-x_2 -v_h \tau_{12})^{S} (-x_1+x_2 -v_h \tau_{12})^{S}} \\
\times&\frac{1}{ (x_1+x_2 -v_h \tau_{12})^{Y} (-x_1-x_2 -v_h \tau_{12})^{Z}(x_1+x_2 -v_F \tau_{12})^{0.5}}\\
\Big\langle T\mbox{  }\psi&_{L}(X_1)\psi_{R}^{\dagger}(X_2)\Big\rangle \sim 
\frac{(2x_1)^{\gamma_1}(2x_2)^{1+\gamma_2}+(2x_1)^{1+\gamma_2}(2x_2)^{\gamma_1}}{2(x_1-x_2 -v_h \tau_{12})^{S} (-x_1+x_2 -v_h \tau_{12})^{S}} \\
\times&\frac{1}{ (x_1+x_2 -v_h \tau_{12})^{Z} (-x_1-x_2 -v_h \tau_{12})^{Y}(-x_1-x_2 -v_F \tau_{12})^{0.5}}\\
\label{SS}
\end{aligned}
\end{equation}

\small
\begin{bf}Case II : $x_1$ and $x_2$ on opposite sides of the origin\end{bf} \\ \scriptsize

\begin{equation}
\begin{aligned}
\Big\langle T\mbox{  }\psi&_{R}(X_1)\psi_{R}^{\dagger}(X_2)\Big\rangle \sim 
\frac{(2x_1)^{1+\gamma_2}(2x_2)^{\gamma_1} }{2(x_1-x_2 -v_h \tau_{12})^{A} (-x_1+x_2 -v_h \tau_{12})^{B}} \\
\times&\frac{(x_1+x_2)^{-1}(x_1+x_2 + v_F \tau_{12})^{0.5}}{ (x_1+x_2 -v_h \tau_{12})^{C} (-x_1-x_2 -v_h \tau_{12})^{D} (x_1-x_2 -v_F \tau_{12})^{0.5}}\\
&\hspace{2cm}+\frac{(2x_1)^{\gamma_1} (2x_2)^{1+\gamma_2}}{2(x_1-x_2 -v_h \tau_{12})^{A} (-x_1+x_2 -v_h \tau_{12})^{B}} \\
\times&\frac{(x_1+x_2)^{-1}(x_1+x_2 - v_F \tau_{12})^{0.5}}{ (x_1+x_2 -v_h \tau_{12})^{D} (-x_1-x_2 -v_h \tau_{12})^{C} (x_1-x_2 -v_F \tau_{12})^{0.5}}\\
\Big\langle T\mbox{  }\psi&_{L}(X_1)\psi_{L}^{\dagger}(X_2)\Big\rangle \sim 
\frac{(2x_1)^{1+\gamma_2}(2x_2)^{\gamma_1} }{2(x_1-x_2 -v_h \tau_{12})^{B} (-x_1+x_2 -v_h \tau_{12})^{A}} \\
\times&\frac{(x_1+x_2)^{-1}(x_1+x_2 - v_F \tau_{12})^{0.5}}{ (x_1+x_2 -v_h \tau_{12})^{D} (-x_1-x_2 -v_h \tau_{12})^{C} (-x_1+x_2 -v_F \tau_{12})^{0.5}}\\
&\hspace{2cm}+\frac{(2x_1)^{\gamma_1} (2x_2)^{1+\gamma_2}}{2(x_1-x_2 -v_h \tau_{12})^{B} (-x_1+x_2 -v_h \tau_{12})^{A}} \\
\times&\frac{(x_1+x_2)^{-1}(x_1+x_2 + v_F \tau_{12})^{0.5}}{ (x_1+x_2 -v_h \tau_{12})^{C} (-x_1-x_2 -v_h \tau_{12})^{D} (-x_1+x_2 -v_F \tau_{12})^{0.5}}\\
\Big\langle T\mbox{  }\psi&_{R}(X_1)\psi_{L}^{\dagger}(X_2)\Big\rangle \sim \mbox{ }0\\
\Big\langle T\mbox{  }\psi&_{L}(X_1)\psi_{R}^{\dagger}(X_2)\Big\rangle \sim  \mbox{ }0\\
\label{OS}
\end{aligned}
\end{equation}
\normalsize
where
\footnotesize
\begin{equation}
Q=\frac{(v_h-v_F)^2}{8 v_h v_F} \mbox{ };\mbox{ }  X=\frac{|R|^2(v_h-v_F)(v_h+v_F)}{8  v_h (v_h-|R|^2 (v_h-v_F))}  \mbox{ };\mbox{ }C=\frac{v_h-v_F}{4v_h}
\label{luttingerexponents}\end{equation}
\normalsize
The other exponents can be expressed in terms of the above exponents.
\footnotesize
\begin{equation*}
\begin{aligned}
&P= \frac{1}{2}+Q  \mbox{ };\hspace{0.8 cm}    S=\frac{Q}{C}( \frac{1}{2}-C)   \mbox{ };\hspace{0.85 cm}      Y=\frac{1}{2}+X-C  ;           \\
& Z=X-C\mbox{ };\hspace{0.8 cm}      A=\frac{1}{2}+Q-X \mbox{ };\hspace{0.8 cm}   B=Q-X  \mbox{ };\hspace{1 cm}   \\
&D=-\frac{1}{2}+C   \mbox{ };\hspace{.6 cm}      \gamma_1=X                \mbox{ };\hspace{1.65 cm}    \gamma_2=-1+X+2C;\\
\end{aligned}
\end{equation*}
\normalsize
\section{Funding}
A part of this work was done with financial support from Department of Science and Technology, Govt. of India DST/SERC: SR/S2/CMP/46 2009.\\


\bibliographystyle{apsrev4-1}
\bibliography{ref}

\begin{thebibliography}{26}%
\makeatletter
\providecommand \@ifxundefined [1]{%
 \@ifx{#1\undefined}
}%
\providecommand \@ifnum [1]{%
 \ifnum #1\expandafter \@firstoftwo
 \else \expandafter \@secondoftwo
 \fi
}%
\providecommand \@ifx [1]{%
 \ifx #1\expandafter \@firstoftwo
 \else \expandafter \@secondoftwo
 \fi
}%
\providecommand \natexlab [1]{#1}%
\providecommand \enquote  [1]{``#1''}%
\providecommand \bibnamefont  [1]{#1}%
\providecommand \bibfnamefont [1]{#1}%
\providecommand \citenamefont [1]{#1}%
\providecommand \href@noop [0]{\@secondoftwo}%
\providecommand \href [0]{\begingroup \@sanitize@url \@href}%
\providecommand \@href[1]{\@@startlink{#1}\@@href}%
\providecommand \@@href[1]{\endgroup#1\@@endlink}%
\providecommand \@sanitize@url [0]{\catcode `\\12\catcode `\$12\catcode
  `\&12\catcode `\#12\catcode `\^12\catcode `\_12\catcode `\%12\relax}%
\providecommand \@@startlink[1]{}%
\providecommand \@@endlink[0]{}%
\providecommand \url  [0]{\begingroup\@sanitize@url \@url }%
\providecommand \@url [1]{\endgroup\@href {#1}{\urlprefix }}%
\providecommand \urlprefix  [0]{URL }%
\providecommand \Eprint [0]{\href }%
\providecommand \doibase [0]{http://dx.doi.org/}%
\providecommand \selectlanguage [0]{\@gobble}%
\providecommand \bibinfo  [0]{\@secondoftwo}%
\providecommand \bibfield  [0]{\@secondoftwo}%
\providecommand \translation [1]{[#1]}%
\providecommand \BibitemOpen [0]{}%
\providecommand \bibitemStop [0]{}%
\providecommand \bibitemNoStop [0]{.\EOS\space}%
\providecommand \EOS [0]{\spacefactor3000\relax}%
\providecommand \BibitemShut  [1]{\csname bibitem#1\endcsname}%
\let\auto@bib@innerbib\@empty
\bibitem [{\citenamefont {Haldane}(1981)}]{haldane1981luttinger}%
  \BibitemOpen
  \bibfield  {author} {\bibinfo {author} {\bibfnamefont {F.}~\bibnamefont
  {Haldane}},\ }\href@noop {} {\bibfield  {journal} {\bibinfo  {journal}
  {Journal of Physics C: Solid State Physics}\ }\textbf {\bibinfo {volume}
  {14}},\ \bibinfo {pages} {2585} (\bibinfo {year} {1981})}\BibitemShut
  {NoStop}%
\bibitem [{\citenamefont {Von~Delft}\ and\ \citenamefont
  {Schoeller}(1998)}]{von1998bosonization}%
  \BibitemOpen
  \bibfield  {author} {\bibinfo {author} {\bibfnamefont {J.}~\bibnamefont
  {Von~Delft}}\ and\ \bibinfo {author} {\bibfnamefont {H.}~\bibnamefont
  {Schoeller}},\ }\href@noop {} {\bibfield  {journal} {\bibinfo  {journal}
  {Annalen der Physik}\ }\textbf {\bibinfo {volume} {7}},\ \bibinfo {pages}
  {225} (\bibinfo {year} {1998})}\BibitemShut {NoStop}%
\bibitem [{\citenamefont {Giamarchi}(2004)}]{giamarchi2004quantum}%
  \BibitemOpen
  \bibfield  {author} {\bibinfo {author} {\bibfnamefont {T.}~\bibnamefont
  {Giamarchi}},\ }\href@noop {} {\emph {\bibinfo {title} {Quantum physics in
  one dimension}}}\ (\bibinfo  {publisher} {Clarendon Oxford},\ \bibinfo {year}
  {2004})\BibitemShut {NoStop}%
\bibitem [{\citenamefont {Matveev}\ \emph {et~al.}(1993)\citenamefont
  {Matveev}, \citenamefont {Yue},\ and\ \citenamefont
  {Glazman}}]{matveev1993tunneling}%
  \BibitemOpen
  \bibfield  {author} {\bibinfo {author} {\bibfnamefont {K.}~\bibnamefont
  {Matveev}}, \bibinfo {author} {\bibfnamefont {D.}~\bibnamefont {Yue}}, \ and\
  \bibinfo {author} {\bibfnamefont {L.}~\bibnamefont {Glazman}},\ }\href@noop
  {} {\bibfield  {journal} {\bibinfo  {journal} {Physical Review Letters}\
  }\textbf {\bibinfo {volume} {71}},\ \bibinfo {pages} {3351} (\bibinfo {year}
  {1993})}\BibitemShut {NoStop}%
\bibitem [{\citenamefont {Das}\ and\ \citenamefont
  {Setlur}(2018{\natexlab{a}})}]{das2018quantum}%
  \BibitemOpen
  \bibfield  {author} {\bibinfo {author} {\bibfnamefont {J.~P.}\ \bibnamefont
  {Das}}\ and\ \bibinfo {author} {\bibfnamefont {G.~S.}\ \bibnamefont
  {Setlur}},\ }\href@noop {} {\bibfield  {journal} {\bibinfo  {journal}
  {International Journal of Modern Physics A}\ ,\ \bibinfo {pages} {1850174}}
  (\bibinfo {year} {2018}{\natexlab{a}})}\BibitemShut {NoStop}%
\bibitem [{\citenamefont {Das}\ and\ \citenamefont
  {Setlur}(2017)}]{das2017one}%
  \BibitemOpen
  \bibfield  {author} {\bibinfo {author} {\bibfnamefont {J.~P.}\ \bibnamefont
  {Das}}\ and\ \bibinfo {author} {\bibfnamefont {G.~S.}\ \bibnamefont
  {Setlur}},\ }\href@noop {} {\bibfield  {journal} {\bibinfo  {journal}
  {Physica E: Low-dimensional Systems and Nanostructures}\ }\textbf {\bibinfo
  {volume} {94}},\ \bibinfo {pages} {216} (\bibinfo {year} {2017})}\BibitemShut
  {NoStop}%
\bibitem [{\citenamefont {Das}\ and\ \citenamefont
  {Setlur}(2018{\natexlab{b}})}]{das2018ponderous}%
  \BibitemOpen
  \bibfield  {author} {\bibinfo {author} {\bibfnamefont {J.~P.}\ \bibnamefont
  {Das}}\ and\ \bibinfo {author} {\bibfnamefont {G.~S.}\ \bibnamefont
  {Setlur}},\ }\href@noop {} {\bibfield  {journal} {\bibinfo  {journal} {EPL
  (Europhysics Letters)}\ }\textbf {\bibinfo {volume} {123}},\ \bibinfo {pages}
  {27002} (\bibinfo {year} {2018}{\natexlab{b}})}\BibitemShut {NoStop}%
\bibitem [{\citenamefont {Egger}\ and\ \citenamefont
  {Grabert}(1995)}]{Egger1995friedel1}%
  \BibitemOpen
  \bibfield  {author} {\bibinfo {author} {\bibfnamefont {R.}~\bibnamefont
  {Egger}}\ and\ \bibinfo {author} {\bibfnamefont {H.}~\bibnamefont
  {Grabert}},\ }\href@noop {} {\bibfield  {journal} {\bibinfo  {journal}
  {Physical Review Letters}\ }\textbf {\bibinfo {volume} {75}},\ \bibinfo
  {pages} {3505} (\bibinfo {year} {1995})}\BibitemShut {NoStop}%
\bibitem [{\citenamefont {Fendley}\ \emph
  {et~al.}(1995{\natexlab{a}})\citenamefont {Fendley}, \citenamefont {Ludwig},\
  and\ \citenamefont {Saleur}}]{fendley1995exact}%
  \BibitemOpen
  \bibfield  {author} {\bibinfo {author} {\bibfnamefont {P.}~\bibnamefont
  {Fendley}}, \bibinfo {author} {\bibfnamefont {A.}~\bibnamefont {Ludwig}}, \
  and\ \bibinfo {author} {\bibfnamefont {H.}~\bibnamefont {Saleur}},\
  }\href@noop {} {\bibfield  {journal} {\bibinfo  {journal} {Physical Review
  Letters}\ }\textbf {\bibinfo {volume} {74}},\ \bibinfo {pages} {3005}
  (\bibinfo {year} {1995}{\natexlab{a}})}\BibitemShut {NoStop}%
\bibitem [{\citenamefont {Fendley}\ \emph
  {et~al.}(1995{\natexlab{b}})\citenamefont {Fendley}, \citenamefont {Ludwig},\
  and\ \citenamefont {Saleur}}]{fendley1995exact2}%
  \BibitemOpen
  \bibfield  {author} {\bibinfo {author} {\bibfnamefont {P.}~\bibnamefont
  {Fendley}}, \bibinfo {author} {\bibfnamefont {A.}~\bibnamefont {Ludwig}}, \
  and\ \bibinfo {author} {\bibfnamefont {H.}~\bibnamefont {Saleur}},\
  }\href@noop {} {\bibfield  {journal} {\bibinfo  {journal} {Physical Review
  B}\ }\textbf {\bibinfo {volume} {52}},\ \bibinfo {pages} {8934} (\bibinfo
  {year} {1995}{\natexlab{b}})}\BibitemShut {NoStop}%
\bibitem [{\citenamefont {Furusaki}\ and\ \citenamefont
  {Nagaosa}(1994)}]{furusaki1994kondo}%
  \BibitemOpen
  \bibfield  {author} {\bibinfo {author} {\bibfnamefont {A.}~\bibnamefont
  {Furusaki}}\ and\ \bibinfo {author} {\bibfnamefont {N.}~\bibnamefont
  {Nagaosa}},\ }\href@noop {} {\bibfield  {journal} {\bibinfo  {journal}
  {Physical Review Letters}\ }\textbf {\bibinfo {volume} {72}},\ \bibinfo
  {pages} {892} (\bibinfo {year} {1994})}\BibitemShut {NoStop}%
\bibitem [{\citenamefont {Schiller}\ and\ \citenamefont
  {Ingersent}(1995)}]{schiller1995exact}%
  \BibitemOpen
  \bibfield  {author} {\bibinfo {author} {\bibfnamefont {A.}~\bibnamefont
  {Schiller}}\ and\ \bibinfo {author} {\bibfnamefont {K.}~\bibnamefont
  {Ingersent}},\ }\href@noop {} {\bibfield  {journal} {\bibinfo  {journal}
  {Physical Review B}\ }\textbf {\bibinfo {volume} {51}},\ \bibinfo {pages}
  {4676} (\bibinfo {year} {1995})}\BibitemShut {NoStop}%
\bibitem [{\citenamefont {Kane}\ and\ \citenamefont
  {Fisher}(1992)}]{kane1992transport}%
  \BibitemOpen
  \bibfield  {author} {\bibinfo {author} {\bibfnamefont {C.}~\bibnamefont
  {Kane}}\ and\ \bibinfo {author} {\bibfnamefont {M.~P.}\ \bibnamefont
  {Fisher}},\ }\href@noop {} {\bibfield  {journal} {\bibinfo  {journal}
  {Physical Review Letters}\ }\textbf {\bibinfo {volume} {68}},\ \bibinfo
  {pages} {1220} (\bibinfo {year} {1992})}\BibitemShut {NoStop}%
\bibitem [{\citenamefont {Friedel}(1958)}]{friedel1958metallic}%
  \BibitemOpen
  \bibfield  {author} {\bibinfo {author} {\bibfnamefont {J.}~\bibnamefont
  {Friedel}},\ }\href@noop {} {\bibfield  {journal} {\bibinfo  {journal} {Il
  Nuovo Cimento (1955-1965)}\ }\textbf {\bibinfo {volume} {7}},\ \bibinfo
  {pages} {287} (\bibinfo {year} {1958})}\BibitemShut {NoStop}%
\bibitem [{\citenamefont {T{\"u}tt{\"o}}\ and\ \citenamefont
  {Zawadowski}(1988)}]{tutto1988theory}%
  \BibitemOpen
  \bibfield  {author} {\bibinfo {author} {\bibfnamefont {I.}~\bibnamefont
  {T{\"u}tt{\"o}}}\ and\ \bibinfo {author} {\bibfnamefont {A.}~\bibnamefont
  {Zawadowski}},\ }\href@noop {} {\bibfield  {journal} {\bibinfo  {journal}
  {Physical Review Letters}\ }\textbf {\bibinfo {volume} {60}},\ \bibinfo
  {pages} {1442} (\bibinfo {year} {1988})}\BibitemShut {NoStop}%
\bibitem [{\citenamefont {Simion}\ and\ \citenamefont
  {Giuliani}(2005)}]{simion2005friedel}%
  \BibitemOpen
  \bibfield  {author} {\bibinfo {author} {\bibfnamefont {G.~E.}\ \bibnamefont
  {Simion}}\ and\ \bibinfo {author} {\bibfnamefont {G.~F.}\ \bibnamefont
  {Giuliani}},\ }\href@noop {} {\bibfield  {journal} {\bibinfo  {journal}
  {Physical Review B}\ }\textbf {\bibinfo {volume} {72}},\ \bibinfo {pages}
  {045127} (\bibinfo {year} {2005})}\BibitemShut {NoStop}%
\bibitem [{\citenamefont {Emery}(1979)}]{emery1979highly}%
  \BibitemOpen
  \bibfield  {author} {\bibinfo {author} {\bibfnamefont {V.}~\bibnamefont
  {Emery}},\ }\href@noop {} {\enquote {\bibinfo {title} {Highly conducting
  one-dimensional solids},}\ } (\bibinfo {year} {1979})\BibitemShut {NoStop}%
\bibitem [{\citenamefont {Saito}(1978)}]{saito1978self}%
  \BibitemOpen
  \bibfield  {author} {\bibinfo {author} {\bibfnamefont {Y.}~\bibnamefont
  {Saito}},\ }\href@noop {} {\bibfield  {journal} {\bibinfo  {journal}
  {Zeitschrift f{\"u}r Physik B Condensed Matter}\ }\textbf {\bibinfo {volume}
  {32}},\ \bibinfo {pages} {75} (\bibinfo {year} {1978})}\BibitemShut {NoStop}%
\bibitem [{\citenamefont {Gogolin}(1993)}]{gogolin1993local}%
  \BibitemOpen
  \bibfield  {author} {\bibinfo {author} {\bibfnamefont {A.~O.}\ \bibnamefont
  {Gogolin}},\ }\href@noop {} {\bibfield  {journal} {\bibinfo  {journal}
  {Physical review letters}\ }\textbf {\bibinfo {volume} {71}},\ \bibinfo
  {pages} {2995} (\bibinfo {year} {1993})}\BibitemShut {NoStop}%
\bibitem [{\citenamefont {Fern{\'a}ndez}\ and\ \citenamefont
  {Na{\'o}n}(2001)}]{fernandez2001friedel}%
  \BibitemOpen
  \bibfield  {author} {\bibinfo {author} {\bibfnamefont {V.~I.}\ \bibnamefont
  {Fern{\'a}ndez}}\ and\ \bibinfo {author} {\bibfnamefont {C.~M.}\ \bibnamefont
  {Na{\'o}n}},\ }\href@noop {} {\bibfield  {journal} {\bibinfo  {journal}
  {Physical Review B}\ }\textbf {\bibinfo {volume} {64}},\ \bibinfo {pages}
  {033402} (\bibinfo {year} {2001})}\BibitemShut {NoStop}%
\bibitem [{\citenamefont {Grishin}\ \emph {et~al.}(2004)\citenamefont
  {Grishin}, \citenamefont {Yurkevich},\ and\ \citenamefont
  {Lerner}}]{grishin2004functional}%
  \BibitemOpen
  \bibfield  {author} {\bibinfo {author} {\bibfnamefont {A.}~\bibnamefont
  {Grishin}}, \bibinfo {author} {\bibfnamefont {I.~V.}\ \bibnamefont
  {Yurkevich}}, \ and\ \bibinfo {author} {\bibfnamefont {I.~V.}\ \bibnamefont
  {Lerner}},\ }\href@noop {} {\bibfield  {journal} {\bibinfo  {journal}
  {Physical Review B}\ }\textbf {\bibinfo {volume} {69}},\ \bibinfo {pages}
  {165108} (\bibinfo {year} {2004})}\BibitemShut {NoStop}%
\bibitem [{\citenamefont {Bockrath}\ \emph {et~al.}(1999)\citenamefont
  {Bockrath}, \citenamefont {Cobden}, \citenamefont {Lu}, \citenamefont
  {Rinzler}, \citenamefont {Smalley}, \citenamefont {Balents},\ and\
  \citenamefont {McEuen}}]{bockrath1999luttinger}%
  \BibitemOpen
  \bibfield  {author} {\bibinfo {author} {\bibfnamefont {M.}~\bibnamefont
  {Bockrath}}, \bibinfo {author} {\bibfnamefont {D.~H.}\ \bibnamefont
  {Cobden}}, \bibinfo {author} {\bibfnamefont {J.}~\bibnamefont {Lu}}, \bibinfo
  {author} {\bibfnamefont {A.~G.}\ \bibnamefont {Rinzler}}, \bibinfo {author}
  {\bibfnamefont {R.~E.}\ \bibnamefont {Smalley}}, \bibinfo {author}
  {\bibfnamefont {L.}~\bibnamefont {Balents}}, \ and\ \bibinfo {author}
  {\bibfnamefont {P.~L.}\ \bibnamefont {McEuen}},\ }\href@noop {} {\bibfield
  {journal} {\bibinfo  {journal} {Nature}\ }\textbf {\bibinfo {volume} {397}},\
  \bibinfo {pages} {598} (\bibinfo {year} {1999})}\BibitemShut {NoStop}%
\bibitem [{\citenamefont {Auslaender}\ \emph {et~al.}(2002)\citenamefont
  {Auslaender}, \citenamefont {Yacoby}, \citenamefont {De~Picciotto},
  \citenamefont {Baldwin}, \citenamefont {Pfeiffer},\ and\ \citenamefont
  {West}}]{auslaender2002tunneling}%
  \BibitemOpen
  \bibfield  {author} {\bibinfo {author} {\bibfnamefont {O.}~\bibnamefont
  {Auslaender}}, \bibinfo {author} {\bibfnamefont {A.}~\bibnamefont {Yacoby}},
  \bibinfo {author} {\bibfnamefont {R.}~\bibnamefont {De~Picciotto}}, \bibinfo
  {author} {\bibfnamefont {K.}~\bibnamefont {Baldwin}}, \bibinfo {author}
  {\bibfnamefont {L.}~\bibnamefont {Pfeiffer}}, \ and\ \bibinfo {author}
  {\bibfnamefont {K.}~\bibnamefont {West}},\ }\href@noop {} {\bibfield
  {journal} {\bibinfo  {journal} {Science}\ }\textbf {\bibinfo {volume}
  {295}},\ \bibinfo {pages} {825} (\bibinfo {year} {2002})}\BibitemShut
  {NoStop}%
\bibitem [{\citenamefont {Stone}(1994)}]{stone1994bosonization}%
  \BibitemOpen
  \bibfield  {author} {\bibinfo {author} {\bibfnamefont {M.}~\bibnamefont
  {Stone}},\ }\href@noop {} {\emph {\bibinfo {title} {Bosonization}}}\
  (\bibinfo  {publisher} {World Scientific},\ \bibinfo {year}
  {1994})\BibitemShut {NoStop}%
\bibitem [{\citenamefont {Fabrizio}\ and\ \citenamefont
  {Gogolin}(1997)}]{fabrizio1997comment}%
  \BibitemOpen
  \bibfield  {author} {\bibinfo {author} {\bibfnamefont {M.}~\bibnamefont
  {Fabrizio}}\ and\ \bibinfo {author} {\bibfnamefont {A.~O.}\ \bibnamefont
  {Gogolin}},\ }\href@noop {} {\bibfield  {journal} {\bibinfo  {journal}
  {Physical review letters}\ }\textbf {\bibinfo {volume} {78}},\ \bibinfo
  {pages} {4527} (\bibinfo {year} {1997})}\BibitemShut {NoStop}%
\bibitem [{\citenamefont {Furusaki}(1997)}]{furusaki1997local}%
  \BibitemOpen
  \bibfield  {author} {\bibinfo {author} {\bibfnamefont {A.}~\bibnamefont
  {Furusaki}},\ }\href@noop {} {\bibfield  {journal} {\bibinfo  {journal}
  {Physical Review B}\ }\textbf {\bibinfo {volume} {56}},\ \bibinfo {pages}
  {9352} (\bibinfo {year} {1997})}\BibitemShut {NoStop}%
\end{thebibliography}%
\normalsize

\end{document}